\documentclass[11pt]{article}
\usepackage{amsmath,amssymb,amsfonts,graphicx,hyperref}
\input{tcilatex}
\begin{document}

\title{Quantum Neural Computation for Option Price Modelling}
\author{Vladimir G. Ivancevic\\{\small Defence Science \& Technology Organisation, Australia}}\date{}
\maketitle

\begin{abstract}
We propose a new cognitive framework for option price modelling, using quantum neural computation formalism. Briefly, when we apply a classical nonlinear neural-network learning to a linear quantum Schr\"odinger equation, as a result we get a nonlinear Schr\"odinger equation (NLS), performing as a quantum stochastic filter. In this paper, we present a bidirectional quantum associative memory model for the Black--Scholes--like option price evolution, consisting of a pair of coupled NLS equations, one governing the stochastic volatility and the other governing the option price, both self-organizing in an adaptive `market heat potential', trained by continuous Hebbian learning. This stiff pair of NLS equations is numerically solved using the method of lines with adaptive step-size integrator.\\

\end{abstract}

%\tableofcontents

\section{Introduction}

The celebrated Black--Scholes partial differential equation (PDE) describes
the time--evolution of the market value of a \textit{stock option} \cite%
{BS,Merton}. Formally, for a function $u=u(t,s)$ defined on the domain $%
0\leq s<\infty ,~0\leq t\leq T$ and describing the market value of a stock
option with the stock (asset) price $s$, the \emph{Black--Scholes PDE} can be
written (using the physicist notation: $\partial _{z}u=\partial u/\partial z$%
) as a diffusion--type equation,\footnote{Recall that this similarity with diffusion equation led Black and Scholes to obtain their option
price formula as the solution of the diffusion equation with the initial and
boundary conditions given by the option contract terms.} which also resembles the backward Fokker--Planck equation\footnote{Recall that the forward Fokker–-Planck equation (also known as the Kolmogorov forward equation, in which the probabilities diffuse outwards as time moves forwards) describes the time evolution of the probability density function $p=p(t,x)$ for the position $x$ of a particle, and can be generalized to other observables as well \cite{Kadanoff}. Its first use was statistical description of Brownian motion of a particle in a fluid. Applied to the option--pricing process $p=p(t,s)$ with \emph{drift} $D_{1}=D_{1}(t,s)$, \emph{diffusion} $D_{2}=D_{2}(t,s)$ and volatility $\sigma^2$, the forward Fokker--Planck equation reads:
\begin{equation*}
\partial _{t}p=\frac{1}{2}\partial _{ss}\left(
D_{2}\sigma^2p\right)-\partial _{s}\left( D_{1}p\right).
\end{equation*}
The corresponding backward Fokker--Planck equation (which is probabilistic diffusion in reverse,
i.e., starting at the final forecasts, the probabilities diffuse outwards as time moves backwards) reads:
\begin{equation*}
\partial _{t}p=-\frac{1}{2}\sigma^2\partial _{ss}\left(
D_{2}p\right)-\partial _{s}\left( D_{1}p\right).
\end{equation*}}
\begin{equation}
\partial _{t}u=-\frac{1}{2}(\sigma s)^{2}\,\partial _{ss}u-rs\,\partial
_{s}u+ru,  \label{BS}
\end{equation}%
where $\sigma >0$ is the standard deviation, or \emph{volatility} of $s$, $r$
is the short--term prevailing continuously--compounded risk--free interest
rate, and $T>0$ is the time to maturity of the stock option. In this
formulation it is assumed that the \emph{underlying} (typically the stock)
follows a \emph{geometric Brownian motion} with `drift' $\mu $ and volatility  $\sigma$, given by the stochastic differential equation
(SDE) \cite{Osborne}
\begin{equation}
ds(t)=\mu s(t)dt+\sigma s(t)dW(t), \label{gbm}
\end{equation}%
where $W$ is the standard Wiener process.\footnote{%
The economic ideas behind the Black--Scholes option pricing theory translated
to the stochastic methods and concepts are as follows (see \cite{Perello}). First, the option price depends on the stock price and this is a random
variable evolving with time. Second, the efficient market hypothesis \cite{Fama,Jensen}, i.e., the market incorporates instantaneously any information
concerning future market evolution, implies that the random term in the
stochastic equation must be delta--correlated. That is: speculative prices
are driven by white noise. It is known that any white noise can be written
as a combination of the derivative of the Wiener process \cite{Wiener} and white shot
noise (see \cite{Gardiner}). In this framework, the Black--Scholes option
pricing method was first based on the geometric Brownian motion \cite%
{BS,Merton}, and it was lately extended to include white shot noise.\par The PDE (\ref{BS}) is usually derived from SDEs describing the geometric Brownian motion (\ref{gbm}), with the solution given by: $$s(t)= s(0){\rm e}^{(\mu- \frac{1}{2} \sigma^2) t+ \sigma W(t)}.$$ In
mathematical finance, derivation is usually performed using It\^{o} lemma \cite{Ito}
(assuming that the underlying asset obeys the It\^{o} SDE), while in
econophysics it is performed using Stratonovich interpretation (assuming
that the underlying asset obeys the Stratonovich SDE \cite{Stratonovich}) \cite{Gardiner,Perello}.}

The solution of the PDE (\ref{BS}) depends on boundary conditions, subject
to a number of interpretations, some requiring minor transformations of the
basic BS equation or its solution.\footnote{The basic equation (\ref{BS}) can be applied
to a number of one-dimensional models of interpretations of prices given to $%
u$, e.g., puts or calls, and to $s$, e.g., stocks or futures, dividends,
etc. In the first (and most important) example, $u(t,s)=c(t,s)$ is a call on
a \emph{European vanilla option} with exercise price $X$ and
maturity at $T$; then the solution to (\ref{BS}) is given by (see, e.g. \cite{Ingber})
\begin{eqnarray*}
c(s,t) &=&sN(d_{1})-X\mathrm{e}^{-r(T-t)}N(d_{2}), \\
d_{1} &=&\frac{\ln (s/X)+(r+\frac{1}{2}\sigma ^{2})(T-t)}{\sigma (T-t)^{1/2}}
, \\
d_{2} &=&\frac{\ln (s/X)+(r-\frac{1}{2}\sigma ^{2})(T-t)}{\sigma (T-t)^{1/2}},\\
\text{where}\quad N(d)&=&\frac{1}{2}[1+{\rm Erf}(d/\sqrt{2})].
\end{eqnarray*}}
In practice, the volatility is the least known parameter in (\ref{BS}), and
its estimation is generally the most important part of pricing options.
Usually, the volatility is given in a yearly basis, baselined to some
standard, e.g., 252 trading days per year, or 360 or 365 calendar days. However, and especially after the 1987 crash, the geometric Brownian motion model and
the BS formula were unable to reproduce the option price data of real markets.\footnote{Recall that Black--Scholes model assumes that the underlying volatility is constant over the life of the derivative, and unaffected by the changes in the price level of the underlying. However, this model cannot explain long-observed features of the implied volatility surface such as \emph{volatility smile} and skew, which indicate that implied volatility does tend to vary with respect to strike price and expiration. By assuming that the volatility of the underlying price is a stochastic process itself, rather than a constant, it becomes possible to model derivatives more accurately.}
As an alternative, models of financial dynamics based on two-dimensional diffusion processes, known as
stochastic volatility (SV) models \cite{Fouque}, are being widely accepted as a reasonable explanation
for many empirical observations collected under the name of `stylized facts' \cite{Cont}. In such
models the volatility, that is, the standard deviation of returns, originally thought to be
a constant, is a random process coupled with the return in a SDE of the form similar to (\ref{gbm}), so that they both form a two-dimensional
diffusion process governed by a pair of Langevin equations \cite{Fouque,Perello08,Masoliver}.

Using the standard \emph{Kolmogorov probability} approach, instead of the market value of an option given by the Black--Scholes equation (\ref{BS}), we could consider the corresponding probability density function (PDF) given by the backward Fokker--Planck equation (see \cite{Gardiner}). Alternatively, we can obtain the same PDF (for the market value of a stock option), using the \emph{quantum--probability} formalism \cite{ComplexDyn,QuLeap}, as a solution to a time--dependent linear \emph{Schr\"odinger equation} for the
evolution of the complex--valued wave $\psi-$function for
which the absolute square, $|\psi|^2,$ is the PDF (see \cite{Voit}).

In this paper we go a step further and propose a novel general quantum--probability based,\footnote{Note that the domain of validity of the `quantum probability' is not
restricted to the microscopic world \cite{Ume93}. There are macroscopic
features of classically behaving systems, which cannot be explained without
recourse to the quantum dynamics. For example, a field theoretic model leads to the
view of the phase transition as a condensation that is comparable to the
formation of fog and rain drops from water vapor. According to such a model,
the production of activity with long--range correlation in complex systems like financial markets takes
place through the mechanism of spontaneous breakdown of symmetry,
which has been shown to describe long--range correlation in
condensed matter physics. The adoption of such approach
enables modelling of a financial markets and its hierarchy of
components as a fully integrated macroscopic
quantum system, namely as a macroscopic system which is a quantum system not
in the trivial sense that it is made, like all existing matter, by quantum
components such as atoms and molecules, but in the sense that some of its
macroscopic properties can best be described with recourse to quantum
dynamics (see \cite{FreVit06} and references therein).} option--pricing model, which is both \emph{nonlinear} (see \cite{Trippi,Rothman,Ammann,HighDyn}) and \emph{adaptive} (see \cite{Tse,Ingber,GeoDyn,ComNonlin}). More precisely, we propose a \emph{quantum neural computation} \cite{IvQNC} approach to option price modelling and simulation. Note that this approach is in spirit of our adaptive path integral \cite{IA,IAY} approach to human cognition.

\section{The Model}

\subsection{Bidirectional, spatio-temporal, complex-valued associative memory machine}

The new model is defined as a self-organized system of two coupled nonlinear Schr\"odinger (NLS) equations:\footnote{NLS can be viewed as a `deformation' of the linear time-dependent Schr\"odinger equation from non-relativistic quantum mechanics. NLS no
longer has a physical interpretation as the evolution of a quantum particle,
but can be derived as a model for quantum media such as Bose--Einstein
condensates (see e.g. \cite{Spohn}). In addition, for some values of parameters and initial/boundary conditions, the PDF--solution to NLS can be a \emph{soliton}. This hypothetical \emph{option--price--soliton} can potentially give a new light on financial markets simulation and will be explored elsewhere.} one defining the \emph{option--price wave function} $\psi=\psi(t,s)$, with the corresponding \emph{option--price PDF} defined by $|\psi(t,s)|^2$, and the other defining the \emph{volatility wave function} $\sigma=\sigma(t,\sigma)$, with the corresponding \emph{volatility PDF} defined by $|\sigma(t,\sigma)|^2$. The two focusing NLS equations are coupled so that the volatility PDF is a parameter in the option--price NLS, while the option--price PDF is a parameter in the volatility NLS. In addition, both processes evolve in a common self--organizing \emph{market heat potential}.

Formally, we propose an adaptive, semi-symmetrically coupled, volatility + option--pricing model (with interest rate $r$, imaginary unit $\mathrm{i}=%
\sqrt{-1}$ and Hebbian learning rate $c$), which represents a bidirectional NLS--based  spatio-temporal associative memory. The model is defined (in natural quantum units) by the following stiff NLS--system:\footnote{The proposed coupled NLS equations (\ref{stochVol}) and (\ref{stochPrice})
are both examples of the standard \textit{focusing} NLS equation (see, e.g. \cite{Tao1}):
\begin{equation}
\mathrm{i}\partial _{t}\psi =-\frac{1}{2}\partial _{xx}\psi +V|\psi
|^{2}\psi ,  \label{nlsForm}
\end{equation}%
where $\psi :\mathbb{R}\times \mathbb{R}\rightarrow \mathbb{C}.$ It is a well-known fact that this
equation has both soliton and multisoliton solutions (see \cite{Hasimoto,Tao2}).
Technically speaking, the NLS equation (\ref{nlsForm}) is called `mass and energy sub--critical' \cite{TaoEnerg},
`scattering--critical' \cite{Ginibre}, `Galilean invariant' and `completely
integrable system' \cite{Cazenave,Kato}, arising from classical Hamiltonian energy function
\begin{equation*}
H=\int dx\left( {\frac{1}{2}}|\partial _{x}\psi |^{2}+{\frac{V}{2}}|\psi
|^{4}\right) ,
\end{equation*}%
with the classical Poisson brackets%
$$
\{\psi (x),\psi (y)\}=\{\psi ^{\ast }(x),\psi ^{\ast }(y)\}=0, \qquad
\{\psi ^{\ast }(x),\psi (y)\}=i\delta (s-y).
$$
Its quantum Hamiltonian operator has the form
\begin{equation*}
H=\int dx\left( {\frac{1}{2}}\partial _{x}\psi ^{\dagger }\partial _{x}\psi +%
{\frac{V}{2}}\psi ^{\dagger }\psi ^{\dagger }\psi \psi \right) ,
\end{equation*}%
and the classical Poisson brackets are replaced by quantum Poisson brackets, or commutators:
$$
[\psi(x),\psi(y)]=[\psi^*(x),\psi^*(y)]=0,\qquad
[\psi^*(x),\psi(y)]=-\delta(x-y).
$$
Linear
component of (\ref{nlsForm}) is the standard Schr\"{o}dinger equation:
\begin{equation*}
\mathrm{i}\partial _{t}\psi =-\frac{1}{2}\partial _{xx}\psi +V\psi.
\end{equation*}
The NLS equation (\ref{nlsForm}) has a `blowup' whenever the Hamiltonian $H$ is negative \cite{Glassey,Ogawa} and in particular one has a blowup arbitrarily close to the ground state \cite{Shatah,Zhang}.}
\begin{eqnarray}
\text{Volatility NLS :}\quad \mathrm{i}\partial _{t}\sigma  &=&-\frac{1}{2}%
s^{2}|\psi |^{2}\partial _{ss}\sigma +V(w)|\sigma |^{2}\sigma,
\label{stochVol} \\
\text{Option price NLS :}\quad \mathrm{i}\partial _{t}\psi  &=&-\frac{1}{2}%
s^{2}|\sigma |^{2}\,\partial _{ss}\psi + |\psi |^{2}\psi + r\psi ,
\label{stochPrice} \\ \text{with :}~~ V(w)&=&\sum_{i=1}^N w_{i}g_{i}\qquad \text{and}\notag \\
\text{Adaptation ODE :}\quad \dot{w}_{i} &=&-w_{i}+ c|\sigma |g_{i}|\psi
|.  \label{Hebb1}
\end{eqnarray}
In the proposed model, the $\sigma$--NLS (\ref{stochVol}) governs the short--range PDF--evolution for stochastic volatility, which plays the role of a nonlinear (variable) coefficient in (\ref{stochPrice}); the $\psi$--NLS (\ref{stochPrice}) defines the long--range PDF--evolution for stock price, which plays the role of a nonlinear coefficient in (\ref{stochVol}). The purpose of this coupling is to generate a \emph{leverage effect}, i.e. stock volatility is (negatively) correlated to stock returns\footnote{The hypothesis that financial leverage can explain the leverage effect was first discussed by F. Black \cite{Bl76}.} (see, e.g. \cite{Roman}). The $w-$ODE (\ref{Hebb1}) defines the $(\sigma,\psi)-$based continuous Hebbian learning \cite{NeuFuz,CompMind}. The adaptive volatility potential $V(w)$ is defined as a scalar product of the weight vector $w_{i}$ and the Gaussian kernel vector $g_{i}$ and can be related to the market \emph{temperature} (which obeys Boltzman distribution \cite{KleinertBk}). The Gaussian vector $g_{i}$ is defined as:
\begin{eqnarray}
&&g_{i}={\rm exp}[-(d - m_i d)^2],\qquad d=y_{target}-y, \qquad \text{with}\\
&&y=2\sin(60t), \qquad y_{target}=s|\sigma |^{2}ds,\qquad (i=1,...,N),
\end{eqnarray}
where $m_i={\rm random}(-1.0, 1.0)$,~ while ~$ds = (s_1-s_0)/(N-1)$ represents the stock--price increment defined using the \emph{method of lines} (where each NLS was decomposed into $N$ first-order ODEs; see Appendix).

Note that each of the NLS equations (\ref{stochVol}) and (\ref{stochPrice}) individually resembles `quantum stochastic filtering/quantum neural network' models of
\cite{BeheraFilt,Behera05,Behera06}. Thus, the whole model effectively performs quantum neural computation \cite{IvQNC}, by giving a spatio-temporal and quantum generalization of Kosko's BAM family of neural networks \cite{Kosko1,Kosko2}. In addition, the solitary nature of NLS equations may describe brain-like effects frequently occurring in financial markets: volatility/price soliton propagation, reflection and collision (see \cite{Han}).

\subsection{Simulation results}

The model (\ref{stochVol})--(\ref{Hebb1}) has been numerically solved for the following initial conditions (IC) and repeatable boundary conditions (BC):
\begin{eqnarray}
{\rm IC}&:& \sigma_i(0,s) = 0.25, \qquad \psi_i(0,s) = 1,\\ {\rm BC}&:& \partial _{t}\sigma(t,s_0) = \partial _{t}\sigma(t,s_1),\qquad \partial _{t}\psi(t,s_0) = \partial _{t}\psi(t,s_1),\end{eqnarray}
using the method of lines \cite{MOL} (with $N=30$ lines per NLS discretization, see Appendix) within the Cash--Karp version \cite{Cash} of the adaptive step-size Runge-Kutta-Fehlberg 4/5th order algorithm \cite{Fehlberg}. The average simulation time (depending on the random initial weights and Gaussians) was 10--30 seconds on a standard Pentium 4 PC, using Visual C++ compiler\footnote{In this respect it is important to note that in order to have a useful option pricing model, the speed of calculation is paramount. Especially for European options it is extremely important to have a fast calculation of the price and the hedge parameters, since these are the most liquidly traded financial options.}
\begin{figure}[htb]
\centering \includegraphics[width=15cm]{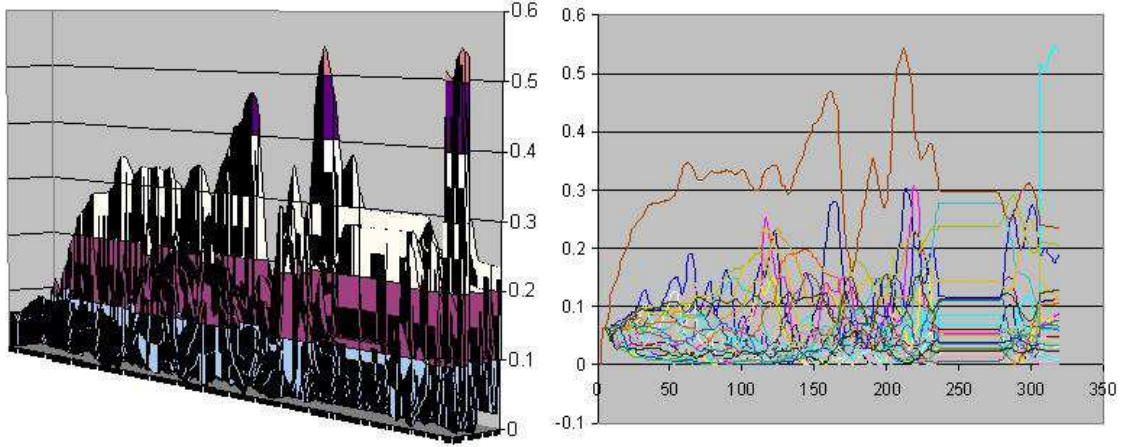}
\caption{A sample daily evolution of the short--range volatility PDF: global surface (left) and individual lines (right).}
\label{VolSurfPDF}
\end{figure}

\begin{figure}[htb]
\centering \includegraphics[width=15cm]{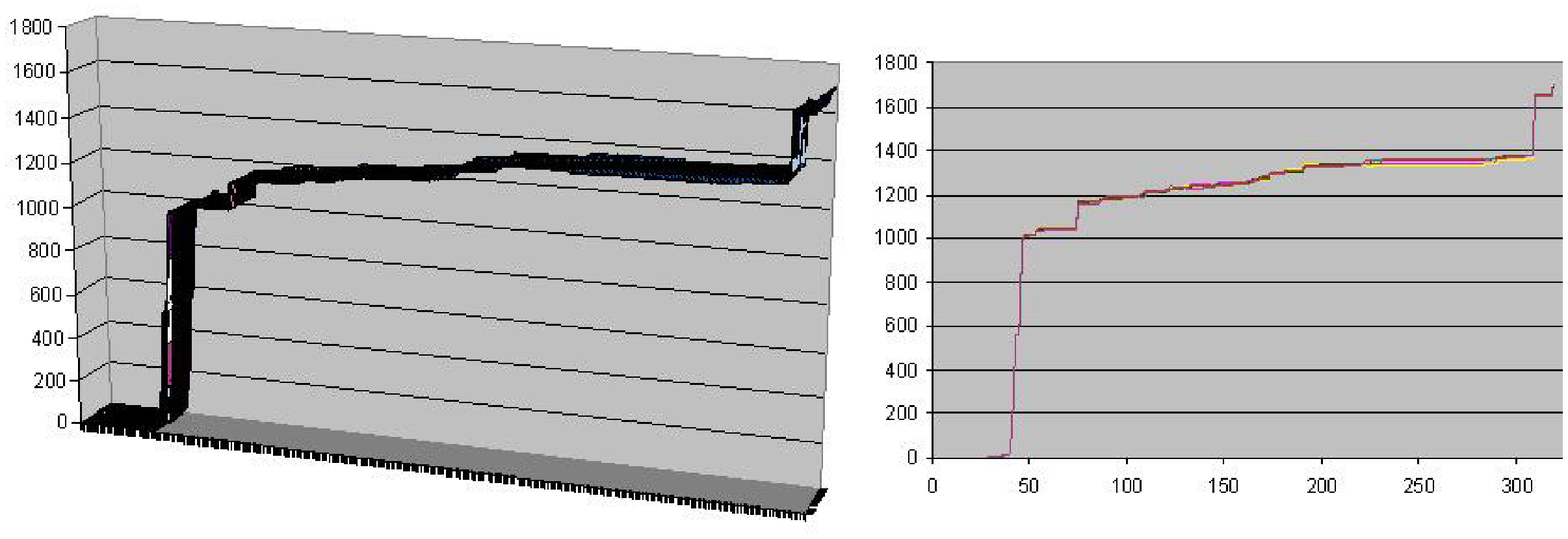}
\caption{A sample daily evolution of the long--range option-price PDF (corresponding to the volatility from Figure \ref{VolSurfPDF}): global surface (left) and individual lines (right).}
\label{PriceSurPDF}
\end{figure}

\begin{figure}[htb]
\centering \includegraphics[width=15cm]{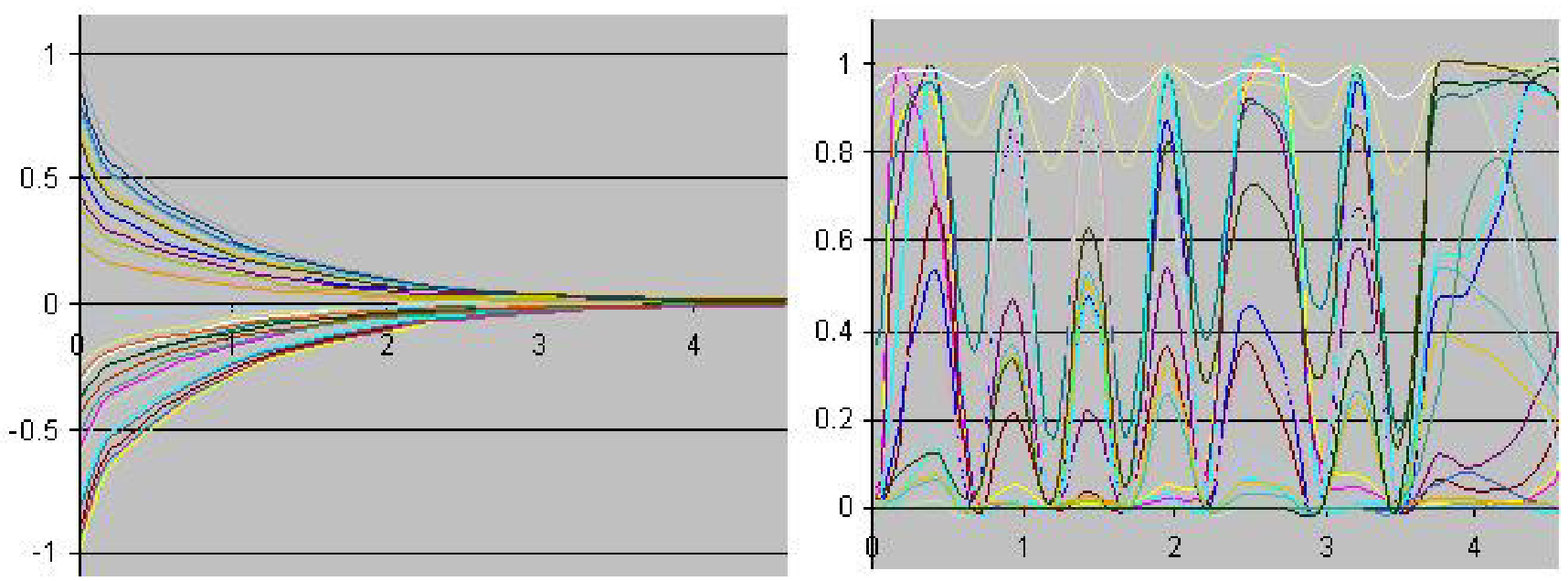}
\caption{A bimonthly evolution of Hebbian weights (left) and Gaussian kernels (right). We can see the stabilizing effect of the Hebbian self-organization (all weights tend to zero).}
\label{Weights}
\end{figure}

A sample daily evolution of the volatility PDF is given in Figure \ref{VolSurfPDF}. The corresponding option-price PDF is given in Figure \ref{PriceSurPDF}, with the Log-plot given in Figure \ref{PriceSurPDF}. The corresponding Hebbian weights and Gaussian kernels are given in Figure \ref{Weights}.

\begin{figure}[htb]
\centering \includegraphics[width=15cm]{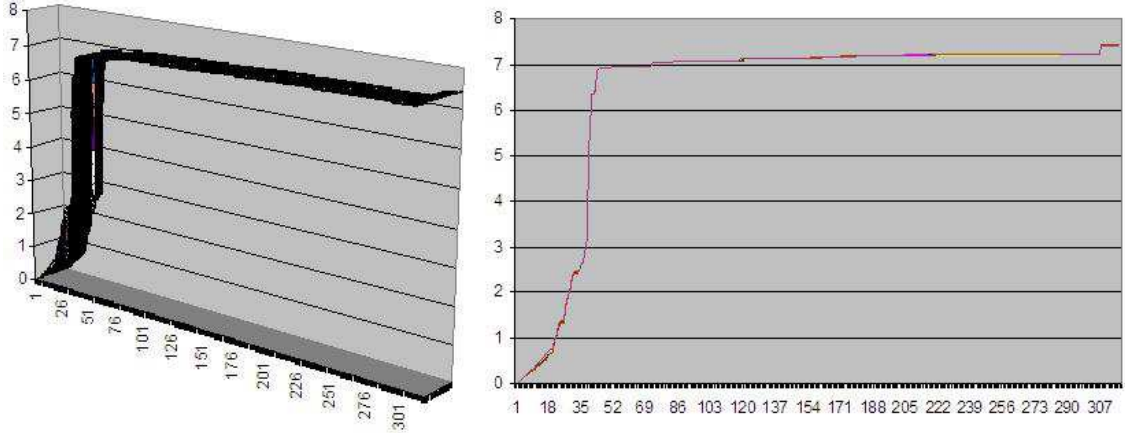}
\caption{Log-plot of the long--range option-price PDF from Figure \ref{PriceSurPDF}.}
\label{LogPlot}
\end{figure}

\subsection{Complex-valued wave-function lines}

\begin{figure}[htb]
\centering \includegraphics[width=15cm]{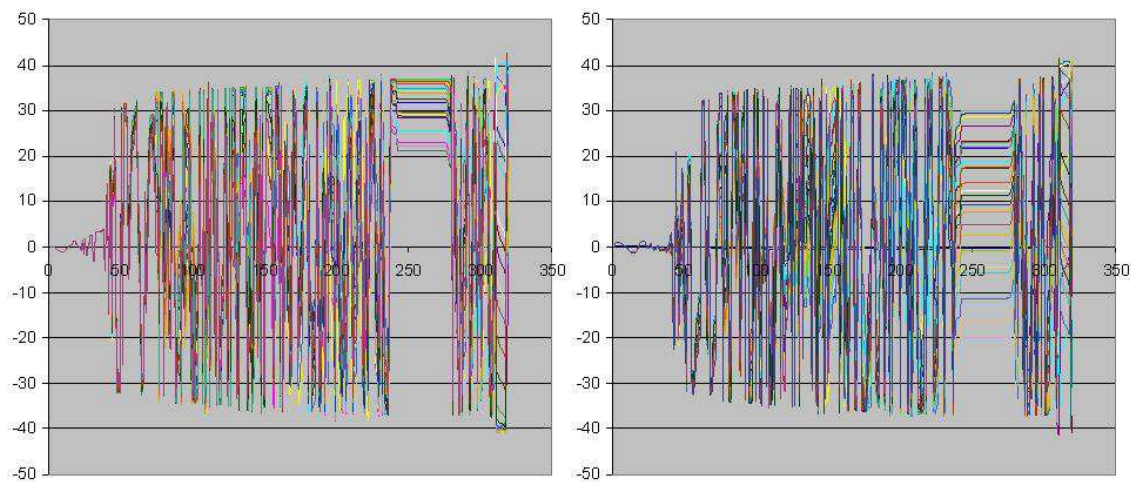}
\caption{Imaginary (left) and real (right) option-price wave-function, corresponding to the PDF from Figure \ref{PriceSurPDF}. Notice: (i) a slow start of the fluctuation, and (ii) the phase transition from 240--240 days from the beginning; this qualitative change of behavior \emph{is not} visible in the PDF.}
\label{ImagReal}
\end{figure}

\begin{figure}[htb]
\centering \includegraphics[width=9cm]{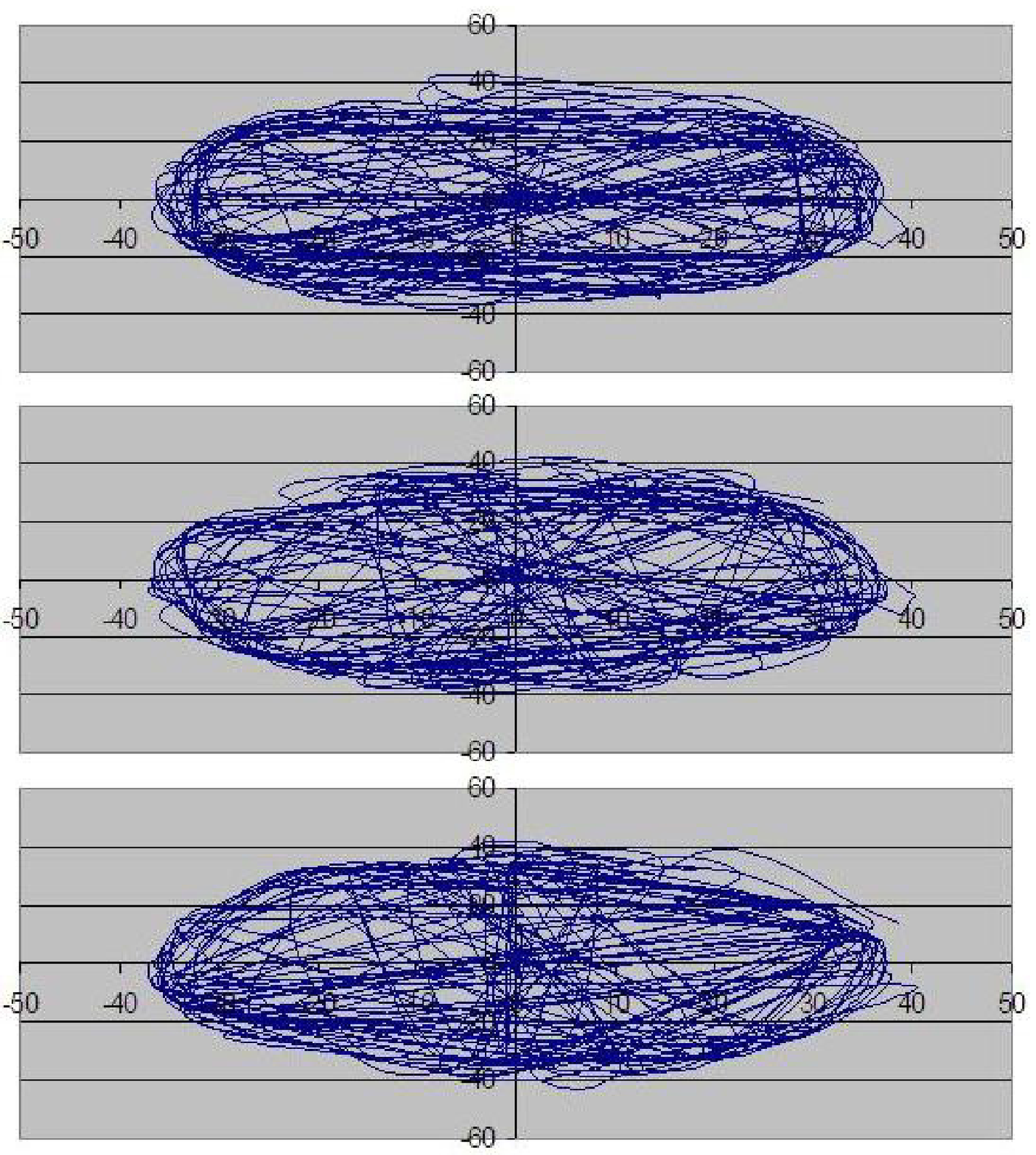}
\caption{Complex-plane phase-plots for three sample wave-function lines corresponding to real versus imaginary option-price lines from Figure \ref{ImagReal}. Notice that all three lines fluctuate around a circular periodic motion.}
\label{PhasePlots}
\end{figure}

Complex-valued quantum probability lines for option price evolution have some interesting properties of their own. Imaginary and real wave-functions corresponding to the option-price PDF from Figure \ref{PriceSurPDF} are given in Figure \ref{ImagReal}. They can be combined in a phase-space fashion to give complex-plane plots, as in Figure \ref{PhasePlots}. These complex phase plots of individual wave-function lines depict small fluctuations around circular periodic motions.

\section{Conclusion}

In this paper we have proposed a new cognitive formalism for option price modelling, based on quantum neural computation. The model consists of a pair of coupled nonlinear Schr\"odinger equations, representing quantum--stochastic volatility and option--price evolutions. The two equations create a leverage effect by self--organizing in a common market--heat potential, using continuous Hebbian learning. An efficient numerical solver for this bidirectional spatio-temporal associative memory is implemented in C++ using the method of lines and adaptive step-size algorithm. Sample simulations are provided as well as the outline of the algorithm.

{\small
\section{Appendix: Method of Lines for Coupled NLS Eqs. (\protect\ref{stochVol})--(\protect\ref{stochPrice})}

Assuming that we already have a stable adaptive Runge-Kutta-Fehlberg (RKF)
integrator for systems of ODEs, we can implement the powerful method of
lines \cite{MOL} for the coupled NLS pair (\ref{stochVol})--(\ref{stochPrice}%
) as follows. We start with the standard heat equation:
\begin{equation}
\partial _{t}\psi =\frac{1}{2}\partial _{xx}\psi ,  \label{heat1}
\end{equation}%
where $\psi =\psi (x,t)$ is a real-valued function. Solution of this PDE can
be found if we have one initial condition:
\begin{equation}
\text{IC}~:~\psi (x,t=0)=\psi _{0}  \label{ic1}
\end{equation}%
and two boundary conditions:
\begin{eqnarray}
\text{BC} ~:&\psi (x=x_{0},t)=\psi _{b},& ~~\text{where }x_{0}\text{ is
the initial value of }x,  \label{bc1} \\
&\partial _{x}\psi (x=x_{f},t)=0,& ~~\text{where }x_{f}\text{ is the
final value of }x.  \notag
\end{eqnarray}

A commonly used second order/central finite difference approximation for $%
\frac{\partial ^{2}\psi }{\partial x^{2}}$ is
\begin{equation*}
\partial _{xx}\psi \approx \frac{\psi _{k+1}-2u_{k}+\psi _{k-1}}{\Delta x^{2}%
},
\end{equation*}%
where $k$ is an index designating a position along a grid in $x$ which has $%
M $ (e.g. 10)\ points\ and $\Delta x$ is the spacing in $x$\ along the grid.
In this way, the heat PDE (\ref{heat1}) can be approximated as a system of
ODEs
\begin{equation}
\frac{d\psi _{k}}{dt}=\frac{1}{2}\frac{\psi _{k+1}-2\psi _{k}+\psi _{k-1}}{%
\Delta x^{2}},\qquad (k=1,2,\ldots N=\text{number of ODEs}).  \label{heat2}
\end{equation}%
The set of ODEs (\ref{heat2}) is then integrated, using the RKF
integrator, subject to IC and BC.

Once we are able accurately solve the set of ODEs (\ref{heat2}) with the
conditions (\ref{ic1})--(\ref{bc1}), we can add the potential term $V(x)$:%
\begin{equation*}
\partial _{t}\psi =\frac{1}{2}\partial _{xx}\psi +V(x)\,\psi ,
\end{equation*}%
with the corresponding set of approximating ODEs:%
\begin{equation}
\frac{d\psi _{k}}{dt}=\frac{1}{2}\frac{\psi _{k+1}-2\psi _{k}+\psi _{k-1}}{%
\Delta x^{2}}+V(k)\,\psi _{k},\qquad (k=1,2,\ldots N).  \label{heat3}
\end{equation}

Once we are able to solve the set of heat ODEs (\ref{heat3}) with the
real-valued conditions (\ref{ic1})--(\ref{bc1}), we can move into the
complex plane, by introducing the imaginary unit $\mathrm{i}=\sqrt{-1}$ on
the left and the minus sign on the right:
\begin{equation}
\mathrm{i}\partial _{t}\psi =-\frac{1}{2}\partial _{xx}\psi +V(x)\,\psi ,
\label{LS1}
\end{equation}%
This is the linear time--dependent Schr\"{o}dinger equation from
non-relativistic quantum mechanics (in natural units, for which both the
Planck constant and the particle mass are one, see \cite{QuLeap}) for the
complex--valued wave $\psi -$function. It has the corresponding set of
approximating complex-valued ODEs:
\begin{equation}
\mathrm{i}\frac{d\psi _{k}}{dt}=-\frac{1}{2}\frac{\psi _{k+1}-2\psi
_{k}+\psi _{k-1}}{\Delta x^{2}}+V(k)\,\psi _{k},\qquad (k=1,2,\ldots N).
\label{sch1}
\end{equation}

Once we are able to solve the linear Schr\"{o}dinger equation (\ref{LS1})
with the complex-valued conditions similar to (\ref{ic1})--(\ref{bc1}), as a
set of complex-valued ODEs (\ref{sch1}), we can replace the linear term on
the right with the cubic nonlinearity:
\begin{equation}
\mathrm{i}\partial _{t}\psi =-\frac{1}{2}\partial _{xx}\psi +V(x)|\psi
|^{2}\psi .
\end{equation}%
This is the nonlinear Schr\"{o}dinger equation (NLS). Its corresponding set
of approximating complex-valued ODEs reads:%
\begin{equation}
\mathrm{i}\frac{d\psi _{k}}{dt}=-\frac{1}{2}\frac{\psi _{k+1}-2\psi
_{k}+\psi _{k-1}}{\Delta x^{2}}+V(x_{k})|\psi |^{2}\psi ,\qquad
(k=1,2,\ldots N).  \label{NLS2}
\end{equation}

Finally, our option pricing model (\ref{stochVol})--(\ref{stochPrice}) represents a stiff pair of coupled complex-valued ODE--decompositions of the form of (\ref{NLS2}), together with the Hebbian learning ODE (\ref{Hebb1}). Although the resulting ODE-system is stiff, the Cash--Karp version \cite{Cash} of the adaptive step-size Runge-Kutta-Fehlberg algorithm \cite{Fehlberg} can efficiently solve it (in Visual C++) for various boundary conditions and random initial weights and Gaussian kernels.
}

\end{document}